\documentstyle[aps,preprint,tighten,floats,epsfig]{revtex}

\def\PR#1{Phys.\ Rev.\ {\bf#1}}
\def\PL#1{Phys.\ Lett.\ {\bf#1}}

\def\be{\begin{equation}}
\def\ee{\end{equation}}
\def\mpi{m_{\pi}}

\begin{document}

\draft
\preprint{\vbox{\noindent\null \hfill ADP-00-18/T402\\ 
                         \null \hfill hep-lat/0004006 \\
}}

\title{Incorporating Chiral Symmetry in Extrapolations of Octet Baryon 
Magnetic Moments}
\author{E.J. Hackett-Jones\footnote{ehackett@physics.adelaide.edu.au}\ , D.B. 
Leinweber\footnote{dleinweb@physics.adelaide.edu.au}\ \ and A.W. 
Thomas\footnote{athomas@physics.adelaide.edu.au}}

\address{Special Research Centre for the Subatomic Structure of 
Matter and}

\address{Department of Physics and Mathematical Physics University of
Adelaide, Australia 5005} 

\maketitle

\begin{abstract}
We explore methods of extrapolating lattice calculations of hadronic 
observables to the physical regime, while respecting the constraints of 
chiral symmetry and heavy quark effective theory. In particular, we 
extrapolate lattice results for magnetic moments of the spin-1/2 baryon 
octet to the physical pion mass and compare with experimental measurements. 
The success previously reported for extrapolations of the nucleon magnetic 
moments carries over to the $\Sigma$ baryons. 
A study of the residual discrepancies in the $\Xi$ baryon 
moments suggests that it is important to have
new simulation data with a more realistic strange quark mass.
\end{abstract}

\newpage

\begin{section}{Introduction}
One of the key goals of lattice QCD is to confront experimental data with the 
predictions of QCD. However, computational limitations mean that hadronic 
observables, such as masses and magnetic moments, are calculated at quark 
masses much larger than their physical values. Although improvements in 
algorithms and computer speed will allow lattice calculations of hadronic 
observables to be performed much closer to the physical regime, these 
improvements will proceed over many years. In the meantime it is imperative 
that one has an understanding of
how to extrapolate lattice results, obtained at large quark masses, to the 
physical world.\vspace{1em} 

A difficult problem encountered in calculating hadronic observables at heavy 
quark masses on the lattice is that chiral perturbation theory is not 
applicable in this heavy quark mass regime. However, chiral perturbation 
theory predicts that near the chiral limit there are important non-analytic 
terms as a function of the quark mass, $m_q$ (or equivalently of $\mpi^2$, as 
$m_{q}\propto\mpi^2$ in this range), in the expansion of a physical 
observable. This non-analytic behaviour must be taken into account in {\it 
any} extrapolation to the physical regime.\vspace{1em}

Here we focus on linking lattice calculations of magnetic moments of the 
spin-$1/2$ baryon octet to the physical world. In particular, we follow an 
extrapolation procedure first introduced for the nucleon magnetic moments 
which builds in the non-analytic behaviour of the magnetic moments near the 
chiral limit, as well as the correct heavy quark behaviour \cite{LLT}. In the 
case of the nucleon, this extrapolation procedure predicts $\mu_p=2.85(22)\ 
\mu_{N}$ and $\mu_{n}=-1.90(15)\ \mu_{N}$ (see Fig.~5 of Ref.\cite{LLT}). This 
agrees well with the experimental measurements of $\mu_{p}=2.793\ \mu_{N}$ and 
$\mu_n=-1.913\ \mu_{N}$. Here we explore the application of this procedure to 
octet baryons in general.\vspace{1em}

The magnetic moment results used here are extracted from the lattice QCD 
calculations of Ref.~\cite{LWD}. While the results are now quite old, they 
continue to be the {\em only} lattice estimates of the spin-$1/2$ baryon octet 
magnetic moments available at the moment. These results were all obtained at 
pion masses above $600$MeV. We extrapolate these results as functions of the 
pion mass, $\mpi$, to the physical pion mass of $140$MeV, to obtain the 
physical magnetic moment predictions. Because the lattice calculations are 
quenched, we expect that there are errors in the lattice data which we have 
been unable to take into account. However, as explained in Ref.~\cite{LLT}, 
these errors are expected to be on the scale of the statistical errors. 
Nevertheless, an ideal extrapolation of magnetic moments would use full QCD 
lattice results which are unavailable at the moment.

\end{section}

\begin{section}{Extrapolations}

To extrapolate the lattice calculations of the magnetic moments we use the 
Pad$\acute{\rm e}$ approximant:
\be
\mu_{i}(\mpi)=\frac{\mu_{0}}{1-\frac{\chi_i}{\mu_{0}}{m_{\pi}}+c\ {m_{\pi}}^2} 
\label{Pade}
\ee
where $\chi_i$, corresponding to the $i^{\rm th}$ baryon, is fixed 
model-independently by chiral perturbation theory and $\mu_{0}$ and $c$ are 
allowed to vary to best fit the data \cite{LLT}. This formula builds in the 
chiral behaviour at small $m_{\pi}$, governed by $\chi_i$, as well as the 
correct heavy quark behaviour, as discussed in the following. \vspace{1em}

The Goldstone boson loops resulting from dynamical chiral symmetry breaking 
mean that the baryon magnetic moments exhibit certain model independent, 
non-analytic behaviour in the quark masses. Using an expansion about the 
chiral SU(3)
limit, one finds that the magnetic moments of the octet 
baryons (in nuclear magnetons, $\mu_N$) are given by
\be 
\mu_i = \gamma_i+\sum_{X=\pi,K}\beta_i^{(X)}\frac{m_N}{8{\pi}f^2}m_{X} +\ldots 
\label{chiral}
\ee
where the ellipses represent higher order terms, including logarithms 
\cite{Jenkins}. Here $f$ is the pion decay constant in the chiral limit ($93$ 
MeV) and $m_N$ is the nucleon mass. For our purposes, namely extrapolating 
lattice data at fixed strange quark mass ($m_s$) as a function of the light 
quark mass ($m_q$), it is preferable to expand about the SU(2) chiral limit. 
The cloudy bag calculations in Ref.\cite{LLT} showed that Goldstone boson 
loops are suppressed like $m_X^{-4}$ at large $m_X$ (comparable to $m_K$). 
Although this result is model dependent, the lattice simulations themselves do 
not show a rapid variation with $m_X$ at values of order $m_K$ or higher, thus 
supporting the general conclusion. One therefore expects that the kaon loops 
should be relatively small and slowly varying as a function of $m_q$ 
\footnote{Recall that $m_K^2 \propto m_s + m_q$ and $m_s$ is fixed and 
large.}. They can therefore be absorbed in the fit parameters $\mu_0$ and $c$. 
On the other hand, the rapid variation of $\mpi$ with $m_q$ means that the 
leading non-analytic behaviour in $\mpi$ must be treated explicitly. 
\vspace{1em}

It is simple to see that the Pad$\acute{\rm e}$ approximant, Eq.~(\ref{Pade}), 
guarantees the correct behaviour of the magnetic moments in the chiral SU(2) 
limit. Expanding Eq.~(\ref{Pade}) about $\mpi=0$ we find
\begin{equation} 
\mu_{i} = \mu_{0} + \chi_{i}\mpi +
\left(\frac{\chi_{i}^2}{\mu_{0}}-\mu_{0}c\right)\mpi^2 + \ldots
\end{equation}
In order to reproduce the leading non-analytic behaviour of the chiral
expansion in our fit we fix $\chi_i$ to the value
$\beta_i^{(\pi)}\left({m_N}/{ 8{\pi}f^2}\right)$ for the $i^{\rm th}$
octet baryon. The one-loop corrected estimates \cite{Jenkins} of the
coefficients $\beta_i^{(\pi)}$ and $\chi_i$ are given in
Table~\ref{table:chi}.\vspace{1em}

\begin{table}[btp] 
\begin{center}
\begin{tabular}{ccccccccc}
&$p$&$n$&$\Lambda$&$\Sigma^{+}$&$\Sigma^{0}$&$\Sigma^{-}$&$\Xi^{0}$&$\Xi^{-}$\\
\hline
$\beta_i^{(\pi)}$&$-(F+D)^2$&$(F+D)^2$&$0$&$-\frac{2}{3}D^2-2F^2$&0&$\frac{2}{3
}D^2+2F^2$&$-(D-F)^2$&$(D-F)^2$\\
$\beta_i^{(\pi)}$&-1.02&1.02&0&-0.57&0&0.57&-0.04&0.04\\
$\chi_i$&-4.41&4.41&0&-2.46&0&2.46&-1.91&1.91\\
\end{tabular}
\end{center}
\caption{One-loop corrected estimates of $\beta_i^{(\pi)}$ (in 
Eq.~(\protect\ref{chiral})) and
$\chi_i=\beta_i^{(\pi)}\left({m_N}/{8{\pi}f^2}\right)$} 
\label{table:chi}
\end{table}

The Pad$\acute{\rm e}$ approximant, Eq. (1), also builds in the expected 
behaviour at large $m_{\pi}$. At heavy quark masses we expect 
that the magnetic moment should fall off as the Dirac moment 
\be
\mu=\frac{e_q}{2m_{q}}\propto\frac{1}{\mpi^2}
\ee 
as $\mpi$ becomes moderately large. This is clearly the case in the 
Pad$\acute{\rm e}$ approximant. Therefore, the Pad$\acute{\rm e}$ approximant 
has been chosen to reproduce physical phenomena at the small and large $\mpi$ 
scales. It also succinctly describes the excellent phenomenology of the Cloudy 
Bag Model
\cite{LLT,CBM}. The Pad$\acute{\rm e}$ approximant has already been used 
successfully in the extrapolation of lattice results of magnetic moments of 
the nucleon, which we include here for completeness \cite{LLT}.\vspace{1em}

\end{section}

\begin{section}{Results} \label{sec:res}
In the following graphs, Figs.~1--4, lattice calculations of the baryon 
magnetic moments are fitted as a function of $m_{\pi}$, according to the 
Pad$\acute{\rm e}$ approximant given in Eq.~(\ref{Pade}), with coefficients, 
$\chi_i$, from Table~\ref{table:chi}. In each case the solid lines are 
Pad$\acute{\rm e}$ approximant fits to the magnetic moment lattice results. 
Experimental measurements are indicated at the physical pion mass by an 
asterisk ($\star$). The magnetic moment predictions of the Pad$\acute{\rm e}$ 
approximant are compared with experimental values in Table~\ref{table:fit}. 
The fit parameters, $\mu_{0}$ and $c$, for the solid lines are also indicated. 
\vspace{1em}

\begin{table}[t]
\begin{center}
\begin{tabular}{cccccc}
\ \ Baryon\ \ & $\ \ \mu_0\ \ $ &$\ \ c\ \ $&Lattice&Averaged 
Lattice&Experiment\\ \hline
$p$&3.46&0.68&2.90(20)& &$2.793$\\ 
$n$&-2.28&0.11&-1.79(21)& &$-1.913$\\
$\Lambda$&-0.38&0.005&-0.38(3)& &$-0.613(4)$\\
$\Sigma^+$&2.71&0.40&2.39(16)&2.53(18)&$2.42(5)$\\
$\Sigma^0$&0.54&0.44&0.53(5)&0.58(7)&0.63(4)\tablenote{The
experimental value for $\Sigma^0$ is taken from the average of
$\Sigma^+$ and $\Sigma^-$ experimental results, which is valid in  
the limit of isospin symmetry.}\\ 
$\Sigma^-$&-1.64&1.35&-1.33(8)&-1.35(15)&$-1.157(25)$\\
$\Xi^0$&-0.80&0.29&-0.82(4)&-0.99(10)&$-1.250(14)$\\ 
$\Xi^-$&-0.46&-0.38&-0.44(2)&-0.67(8)&$-0.69(4)$
\end{tabular}
\end{center}
\caption{Magnetic moments of the octet baryons (in nuclear magnetons) 
predicted by lattice QCD compared with experiment.
The fit parameters $\mu_0$ and $c$ of the Pad$\acute{\rm e}$ approximant are 
also indicated in units of $\mu_N$ and ${\rm GeV}^{-2}$ respectively. The 
column entitled ``Averaged Lattice'' reports magnetic moments from 
extrapolations of lattice calculations averaged to better describe the strange 
quark mass, as discussed in the text.}
\label{table:fit}
\end{table}

In the case of the nucleon, the fits given here (Figs.~1 and 2) are slightly 
different from those given in Ref.~\cite{LLT}, as we omit the second set of 
lattice results (these were extracted from Ref.~\cite{WDL} which dealt with the
nucleon only) in order to produce a consistent set of graphs for the entire 
baryon octet. However, the nucleon fits shown here still give excellent 
agreement with experimental data. The physical magnetic moment predictions for
the $\Sigma^+$ and $\Sigma^-$ are also in good agreement with 
experiment. \vspace{1em}

\begin{figure}[p]
\begin{center}
{\epsfig{file=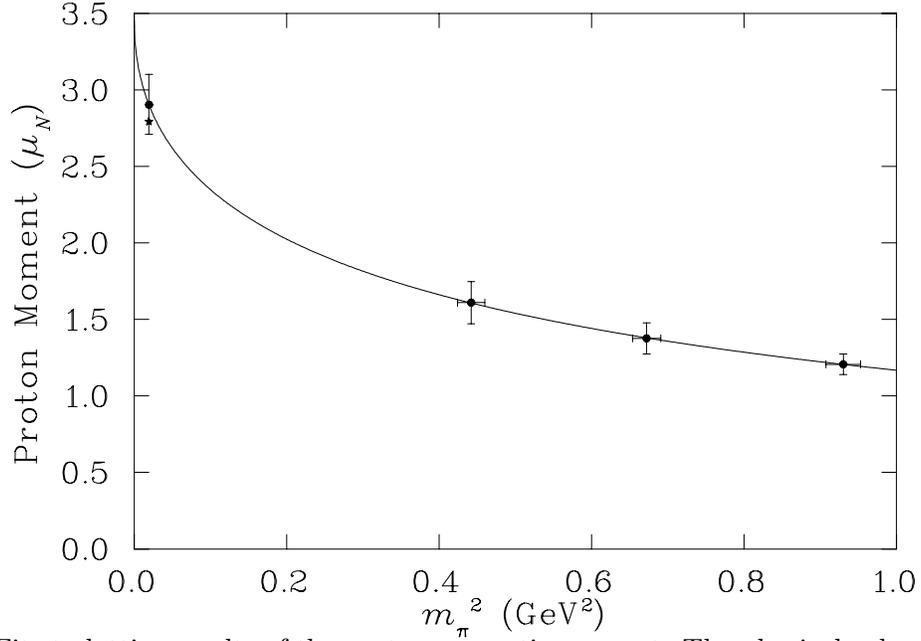, height=12cm, angle=90}}
\caption{Fits to lattice results of the proton magnetic moment. The physical 
value predicted by the fit is also indicated, as is the experimental value, 
denoted by an asterisk.}
\label{fig:prot}
\end{center}
\end{figure}

\begin{figure}[p]
\begin{center}
{\epsfig{file=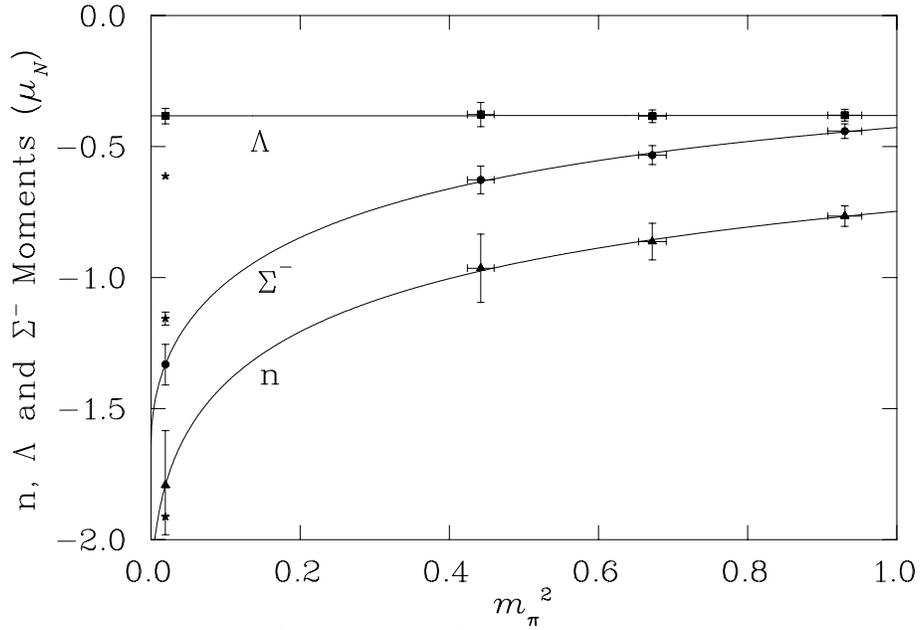, height=12cm, angle=90}}
\caption{Fits to lattice results of the neutron, $\Lambda$ and $\Sigma^-$ 
magnetic moments. The physical values predicted by the fits are indicated, as 
are the experimental values, which are denoted by asterisks.}
\label{fig:siglam}
\end{center}
\end{figure} 

\begin{figure}[t]
\begin{center}
{\epsfig{file=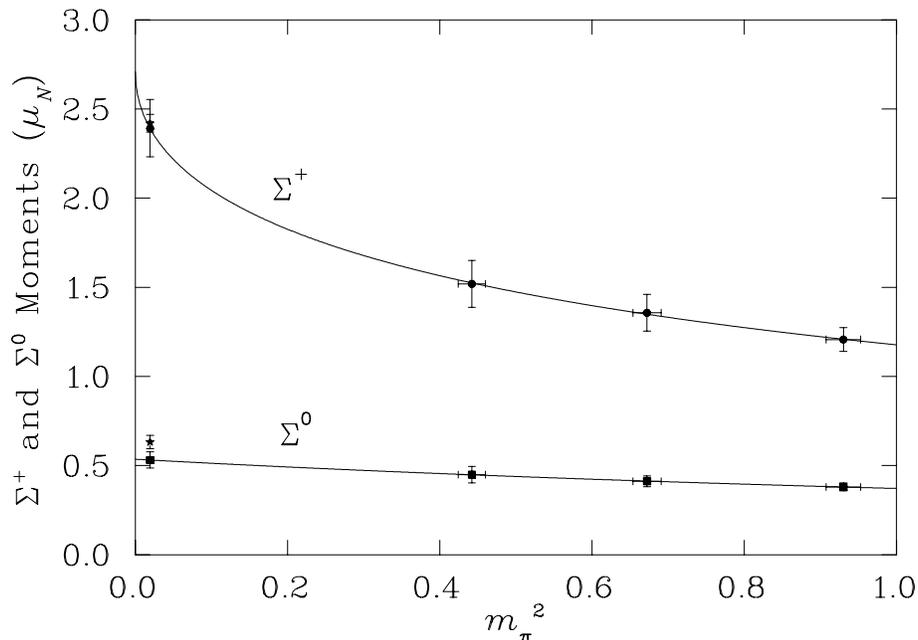, height=12cm, angle=90}}
\caption{Fits to lattice calculations of the $\Sigma^+$ and $\Sigma^0$ 
magnetic moments. The physical values predicted by the fits are indicated, as 
are the experimental values (see text), which are denoted by asterisks.}
\label{fig:sigs}
\end{center}
\end{figure}

Using magnetic moment values predicted by the Pad$\acute{\rm e}$ approximant 
we can calculate the ratio of the $\Xi^-$ and ${\Lambda}$ magnetic moments. 
The simple quark model predicts that this ratio is given by
\be
\frac{\mu_{\Xi^-}}{\mu_{\Lambda}}=\frac{1}{3}
\left(4-\frac{\mu_{d}}{\mu_{s}}\right)
\ee
which becomes 
\be
\frac{\mu_{\Xi^-}}{\mu_{\Lambda}}=\frac{1}{3}\left(4-\frac{m_s}{m_d}\right) 
\ee
if we take each quark magnetic moment to be given by the Dirac moment of its 
constituent mass. In this case the ratio is less than 1 for $m_s > m_d$. This 
disagrees with the experimentally measured value of $1.13(7)$. However, using 
the predictions of the Pad$\acute{\rm e}$ approximant, we obtain a value of 
1.15 for this ratio, which is in excellent agreement with the experimental 
data. This is a good indication that meson cloud effects must be included in 
an extrapolation of lattice results to the physical regime.\vspace{1em}
 
\begin{figure}[t]
\begin{center}
{\epsfig{file=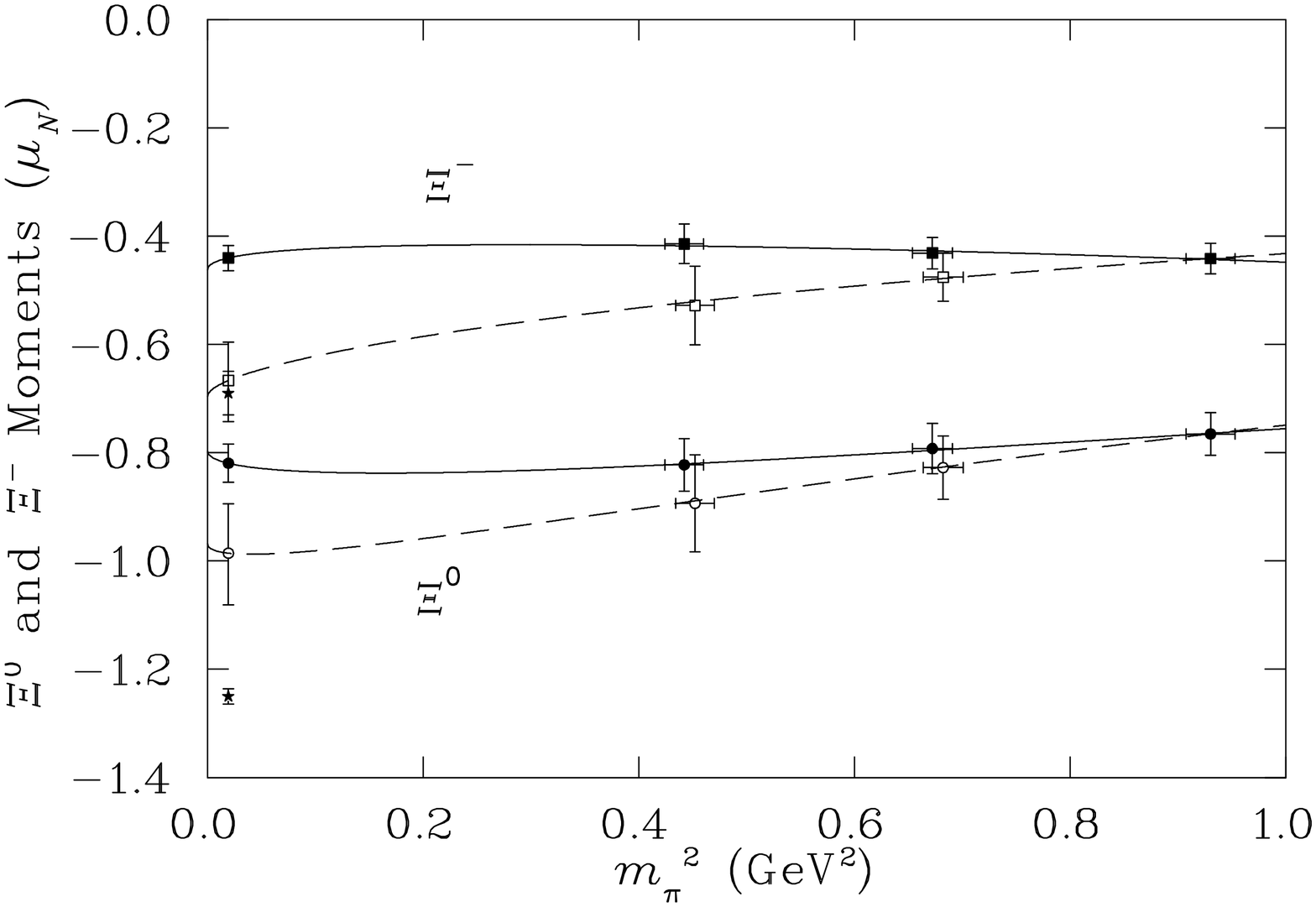, height=12cm, angle=90}}
\caption{Fits to lattice calculations of the $\Xi^0$ and $\Xi^-$ magnetic 
moments. The upper two lines are fits for $\Xi^-$ results and the lower two 
lines for $\Xi^0$ results. Solid lines represent fits to the magnetic moment 
results, whereas dashed lines represent fits to averaged results (denoted by 
open symbols which are offset for clarity), as described in the text. The 
physical values predicted by the fits are indicated, as are the experimental 
values, which are denoted by asterisks.}
\label{fig:xi}
\end{center}
\end{figure}

The lattice calculations of baryon magnetic moments used in this
letter were made with a strange quark mass of approximately $250$\ MeV
\cite{LWD}.  This is much heavier than the physical mass of the
strange quark of $115\pm 8$ MeV at a scale 2 GeV, taken from a careful
analysis of QCD sum rules for $\tau$ decay \cite{Maltman}. The
contribution of the strange quark to the $\Sigma$ baryon magnetic
moments is very small. Lattice QCD calculations indicate that the
contribution of a singly represented quark in a baryon is half that
anticipated by SU(6) spin-flavour symmetry \cite{LA}.  Hence the heavy
strange quark mass will have a subtle effect on the $\Sigma$
moments. By contrast, the strange quarks dominate the $\Lambda$ and
$\Xi$ magnetic moments. Thus the heavy strange quark produces a large
error in the lattice data for these baryons, which so far we have not
taken into account.  This is reflected in the predictions of the
$\Lambda\ $, $\Xi^0$ and $\Xi^-$ magnetic moments which are smaller in
magnitude than the experimental measurements in all
cases. \vspace{1em}

In an attempt to correct for the effect of the large strange quark
mass considered in the lattice calculations, we average the magnetic
moment lattice results of each $S\neq 0$ baryon with magnetic moment
results of a light-quark equivalent baryon \footnote{Lattice
calculations of a light-quark equivalent $\Lambda$ are not
available.}. This procedure interpolates between magnetic moment
lattice results produced with heavy strange quarks and those produced
with zero strange quark mass. These averaged results have an effective
strange quark mass closer to the physical strange quark mass. We have
also used the Pad$\acute{\rm e}$ approximant to extrapolate the
averaged results.  The effect on the $\Sigma$ moments is subtle (see
Table~\ref{table:fit}).  However, in the case of $\Xi^-$, this method
is sufficient to reproduce the empirical $\Xi^-$ moment (as shown by
the dashed line in Fig.~4).  There is a remaining discrepancy in the
value predicted for the $\Xi^0$.  Clearly the present estimate of the
correction for the heavy strange quark mass is somewhat crude.  We
therefore regard it as very important to have new simulation data with
a realistic strange quark mass.  At that stage it may also be
necessary to include kaon loop effects, because the transition $\Xi^0
\rightarrow \Sigma^+ + K^-$ is energetically favoured, and will make a
negative contribution to the $\Xi^0$ magnetic moment.

\end{section}

\begin{section}{Conclusion}
We have shown that the Pad$\acute{\rm e}$ approximant which was
introduced to extrapolate lattice results for the magnetic moments of
the nucleon, is also successful in predicting magnetic moments for the
spin-$1/2$ baryon octet. The magnetic moment values predicted by the
fits for the $p$, $n$, $\Sigma^+$ and $\Sigma^-$ compare well with
experimental data. As a first estimate of the correction to be
expected if a more realistic strange quark mass were used, we averaged
lattice results for the $S\neq{0}$ baryons with the magnetic moments
of the corresponding light-quark baryons. This had a small effect on
the predictions for the $\Sigma$ baryon magnetic moments, but
significantly improved the $\Xi$ baryon results.  In the case of
$\Xi^-$, the averaging procedure produced good agreement with the
experimental results. In the future we hope to perform a similar
extrapolation procedure using more precise magnetic moment lattice
data, calculated with realistic strange quark masses.  At that stage
it may also be necessary to include the kaon loop corrections,
especially for the doubly strange $\Xi$ hyperons.

\end{section}

\section*{Acknowledgement}

This work was supported by the Australian Research Council and the 
University of Adelaide.

\end{document}